\documentstyle{mn}
\def\Zacs{Za\v{c}s}

\def\vv{{\bf v}}

\def\GG{{\bf G}}
\def\PP{{\bf P}}
\def\Teff{$T_{\rm eff}$}
\def\Msun{{\rm M}_\odot}
\def\Rsun{{\rm R}_\odot}
\def\kms{\rm km~s$^{-1}$}
\def\F{{\bf F}}
\def\lan{\langle}
\def\ran{\rangle}
\def\del{\partial }
\def\MBH{$\dot M_{BH}$}

\include{psfig}
\begin{document}
\title{Wind accretion in binary stars -- II. Accretion rates}
\author[T. Theuns, H.M.J. Boffin, A. Jorissen]
	{Tom Theuns$^1$, Henri M.J. Boffin$^{2,3}$, Alain Jorissen$^2$\thanks{
         Research Associate, National Fund for Scientific Research (FNRS), 
         Belgium}\\
 1: Department of Physics, Nuclear Physics Laboratory, Keble Road,
Oxford OX1 3RH, UK\\
 2: Universit\'e Libre de Bruxelles, Institut d'Astronomie et 
d'Astrophysique, Campus Plaine CP 226, Boulevard du
Triomphe, B-1050 Bruxelles\\
 3: Present address : Department of Earth and Planetary Sciences, 
Kobe University, Kobe 657, Japan}

\date{\today}
\maketitle

\begin{abstract}
Smoothed particle hydrodynamics (SPH) is used to estimate accretion
rates of mass, linear and angular momentum in a binary system where
one component undergoes mass loss through a wind. Physical parameters
are chosen such as to model the alleged binary precursors of barium
stars, whose chemical peculiarities are believed to result from the
accretion of the wind from a companion formerly on the asymptotic
giant branch (AGB). The binary system modelled consists of a $3 \Msun$
AGB star (losing mass at a rate $10^{-6}~\Msun~{\rm y}^{-1}$) and a
$1.5 \Msun$ star on the main sequence, in a 3~AU circular orbit.
Three-dimensional simulations are performed for gases with polytropic
indices $\gamma=1$, 1.1 and 1.5, to bracket more realistic situations
that would include radiative cooling. Mass accretion rates are found
to depend on resolution and we estimate typical values of 1-2\% for
the $\gamma=1.5$ case and 8\% for the other models. The highest
resolution obtained (with 400k particles) corresponds to an accretor
of linear size $\approx 16\Rsun$. Despite being (in the $\gamma = 1.5$
case) about ten times smaller than theoretical estimates based on the
Bondi-Hoyle prescription, the SPH accretion rates remain large enough
to explain the pollution of barium stars. Uncertainties in the current
SPH rates remain however, due to the simplified treatment of the wind
acceleration mechanism, as well as to the absence of any cooling
prescription and to the limited numerical resolution.

Angular momentum transfer leads to significant spin up of the accretor
and can account for the rapid rotation of HD~165141, a barium star
with a young white dwarf companion and a rotation rate unusually large
among K giants.

In the circular orbit modelled in this paper, hydrodynamic thrust and
gravitational drag almost exactly compensate and so the net transfer
of linear momentum is nearly zero. For small but finite eccentricities
and the chosen set of parameters, the eccentricity tends to decrease.
\end{abstract}


\section{Introduction}
\label{Sect:Intro}

Wind accretion in a binary system influences the surface composition
and spin of the accreting star and can change the orbital parameters
of the binary.  This paper aims at estimating these effects in a
binary system in which one of the components is an asymptotic giant
branch (AGB) star which suffers from strong mass loss ($\dot M \approx
10^{-6}~\Msun~{\rm y}^{-1}$) through a wind.  The companion main
sequence (MS) star accretes some of this carbon and s-process element
enriched gas and mixes it in its envelope. Stellar evolution will
eventually transform the AGB star into a white dwarf (WD) and the
initially less massive MS star into a chemically peculiar giant.

This is the evolutionary scenario proposed by Boffin \& Jorissen
(1988) for the formation of barium stars, i.e. late-type giants
exhibiting overabundances in carbon and s-process elements
(e.g. Lambert 1985).  All barium stars are likely members of a binary
system with a WD companion (McClure \& Woodsworth 1990, Jorissen \&
Boffin 1992).

The theoretical description of the accretion process is complex due to
the intricate structure of the flow and the complex non-linear
hydrodynamics of the shocked wind material. One can estimate the
expected fraction $\dot M_{acc}/\dot M$ of material being accreted
(Boffin \& Jorissen 1988) from an interpolation formula due to Bondi
\& Hoyle (1944) and Bondi (1952) for the accretion rate onto a star
moving at constant velocity through a gas of uniform density and
temperature. This involves the interpolation of the estimated
accretion rates from two even more simplified models. In the first
model (Hoyle \& Lyttleton 1939), one neglects gas pressure and
estimates the accretion rate $\dot M_{HL}$ onto a compact gravitating
object with mass $M$, moving at constant velocity $v_\infty$ through a
pressureless medium with initially uniform density $\rho_\infty$. In
this model, gravitationally deflected material which passed on one
side of the star collides with material passing on the opposite side,
thereby cancelling its transverse velocity and forming an \lq
accretion line\rq~ behind the star. The material from this accretion
line that has a velocity below the local escape velocity from the star
will be accreted and one finds (Hoyle \& Lyttleton 1939):
\begin{equation}
\dot M_{HL} = \pi R_A^2 \rho_\infty v_\infty,
\label{eq:HL}
\end{equation}
where the accretion radius $R_A$ is given by
\begin{equation}
R_A = 2GM/v^2_\infty.
\label{eq:RA}
\end{equation}
The second model (Bondi 1952; see also Theuns \& David 1992
for the closed form solution of the flow pattern) includes gas
pressure but neglects instead the motion of the star through the
medium. The accretion rate in a steady state is not uniquely
determined but the model does provide a maximum accretion rate $\dot
M_{B}$:
\begin{equation}
\dot M_{B} = \alpha \pi R_B^2 \rho_\infty c_\infty,
\label{eq:B}
\end{equation}
where the Bondi radius is given by Eq.~(\ref{eq:RA}) but with the
sound speed at infinity $c_\infty$ replacing $v_\infty$. The
efficiency parameter $\alpha$ is of the order of unity and 
depends on the value $\gamma$ of the
polytropic index of the gas. The accretion rate $\dot M_{BH}$ for the
Bondi-Hoyle case, where both the velocity of the star as well as gas
pressure are included, is based on an interpolation between these two
extreme cases:
\begin{equation}
\dot M_{BH} = \alpha \pi R_{A}^2 \rho_\infty v_\infty 
              \left(\frac{{\cal M}_\infty^2} 
                    {1 +  {\cal M}_\infty^2}\right)^{3/2},
\label{Eq:BH}
\end{equation}
where ${\cal M}_\infty \equiv v_\infty/c_\infty$ is the Mach number.

The theoretically predicted accretion rate $\dot M_{BH}$ in the
Bondi-Hoyle model has been extensively tested against numerical
simulations by e.g. Hunt (1971) and more recently by Ruffert (1994)
and Ruffert \& Arnett (1994). The theoretical and numerical rates
agree to within 10\% which must be considered a bit fortuitous since
e.g., although transiently present, no accretion line (as envisaged in
the Hoyle and Lyttleton picture) actually exists. The latter
simulations also show the importance of numerical resolution on the
computed accretion rates.

In a previous paper (Theuns \& Jorissen 1992, Paper~I), we presented
three-dimensional (3D) smoothed particle hydrodynamics (SPH)
simulations of wind accretion, taking into account the binary
motion. Two-dimensional simulations of this problem were previously
done by S\o rensen et al. (1975) and by Matsuda et al.  (1987,
1992). Even a glancing comparison between the structure of the flow in
this case (Paper~I) against that in the Bondi-Hoyle model (as in
e.g. Ruffert \& Arnett 1994) shows that the binary motion induces a
very different flow pattern. Consequently, using the accretion rate
\MBH\ in this situation as well might be stretching one's luck too
far.

The main aim of this paper is to test whether the mass accretion rates
are large enough to produce barium stars through the wind accretion
scenario discussed above. The impact of wind accretion on the orbital
parameters and on the spin velocity of the accretor will also be
evaluated. An especially important diagnostic of wind accretion is
provided by the variation of the orbital eccentricity caused by the
transfers of linear momentum between the gas and the accreting
star. Moreover, if the accreted gas possesses angular momentum, the
spin period of the accreting star can be altered. Several aspects
considered in the present paper may also be relevant to the study of
wind accretion in symbiotic stars or in $\zeta$ Aurigae systems.

In addition to the complications introduced by the binary motion,
uncertainties about the thermodynamic state of the gas call for
caution when deducing the accretion rate from oversimplified models.
In fact, the simple polytropic equation of state adopted in this paper
for the gas may not be appropriate in these systems exhibiting \lq
cooling radiation\rq~ in the form of UV light.  Interacting systems
such as symbiotic or extrinsic S stars, where wind accretion is
currently taking place, indeed exhibit a wealth of UV emission lines
or continuum UV radiation (e.g. Nussbaumer \& Stencel 1987; Johnson \&
Ameen 1991). In the interacting S star HD~35155 for example, a total
UV continuum emission of about 0.2~L$_\odot$ is observed (Johnson \&
Ameen 1991). In Paper~I, we identified regions of shocked gas as being
responsible for this kind of emission.

Comparison of the accretion rates obtained for different $\gamma$'s
nevertheless allows us to appraise the importance of the assumed
polytropic index on the computed rates.

\section{Modeling wind accretion with SPH}
\subsection{Description of the numerical method}
\label{Sect:numerical}
All numerical details can be found in Paper~I. Only a short summary
will be given here. Calculations are done in an inertial frame in
which the motion of the centers of mass of the two stars, AGB and MS,
are computed from the solution of the two-body problem for a circular
orbit, neglecting all hydrodynamic effects. The hydrodynamics of the
gas, on the other hand, is treated by means of variable resolution SPH
(Lucy 1977; Gingold \& Monaghan 1977; see e.g. Benz 1989 for a
review).

Gas forces include the hydrodynamic pressure force, $\nabla p/\rho$,
the gravitational force from the MS star and from the AGB star. In
addition, a mechanism responsible for the AGB wind acceleration is
included, simulating the effects of the AGB pulsations and radiation
pressure on the escaping gas (see e.g. Bowen 1988). Since the
luminosity force per unit mass falls with distance like $1/r^2$, just as the
gravitational force does, one can implement such an
acceleration mechanism by decreasing the effective AGB star mass
where it enters the interaction with the gas.  In these simulations,
we assume an accelerating force which exactly balances the
gravitational force from the AGB (resulting in a wind velocity
independent of the distance from the mass-losing star, as obtained in
Bowen's wind models). Self-gravity of the gas is neglected.

A specific model is characterized by the orbital elements of the
binary (radius $A$ of the circular orbit and period $P$ corresponding
to a circular velocity $V_c\equiv 2\pi A/P$), the mass ratio of the two
stars and the properties of the wind, i.e. its polytropic index
$\gamma$ and its velocity, mass flux and sound speed at the surface of
the AGB star.

In addition to these physical parameters, there are additional purely
numerical parameters, most critically the effective resolution. The
latter is determined by the total number of SPH particles used. We
estimate this resolution by computing the average smoothing length
$\overline h$ (see Paper~I) for particles around the accreting
star. That length characterizes the size of the interpolation region
over which quantities are smoothed in the SPH formalism.

The simulations are started with vacuum initial conditions. A
supersonic wind is started around the AGB star and the model is
integrated until a quasi steady-state is reached. The resolution is
then improved by increasing the number of particles emanating from the
AGB star per unit time (yet keeping the mass flux fixed). It is
possible to monitor in this way how the accretion rates change as a
function of resolution.  The accretion process is simulated by
removing mass from the flow, once it enters a sphere with radius
$\overline{h}$ around the accreting star. The average resolution
$\overline{h}$ is computed from all particles in a sphere of radius
$r_h$ around the accreting star.

\subsection{Calculation of fluxes}
\label{Sect:fluxes}
 
Accretion rates are computed by measuring fluxes round the secondary.
Two different methods are used to compute fluxes (see also
Paper~I). The flux $\Phi_{\cal A}$ of a quantity ${\cal A}$ is defined
by
\begin{equation}
\label{Eq:rhoAv}
\Phi_{\cal A} = -\int_\Sigma {\cal A} \rho\vv\cdot d{\bf S},
\end{equation}
where $\rho$ and $\vv$ denote the fluid density and velocity
respectively, $d{\bf S}$ is the outward surface element of a sphere
with radius $R$ centered on the accreting star. In the first method
for evaluating fluxes, the flux density ${\cal A}\rho\vv$ entering
Eq.~(\ref{Eq:rhoAv}) has been evaluated using its SPH estimate on
20x20 points uniformly spaced in $\cos\theta$ and $\phi$ on the
surface of the sphere of radius $R$ (with $\theta$ and $\phi$ being
the usual spherical coordinates with respect to the centre of the
sphere). A numerical integration over the surface of the sphere then
gives $\Phi_{\cal A}$.

Alternatively, $\Phi_{\cal A}$ can be computed by summing over SPH
particles of mass $m_i$:
\begin{equation}
\Phi_{\cal A} = \sum_i {dm_i\over dt} {\cal A}_i.
\label{Eq:dmA}
\end{equation}

For ${\cal A}=1$, $\Phi_{\cal A}$ represents the mass accretion rate.
In that case, $\Phi_{\cal A}$ should be independent of $R$ for a
stationary flow (as long as the sphere does not reach other sources or
sinks, like e.g. the mass-losing star or the outer boundary). The
comparison of the mass accretion rates computed from the flux through
spheres of different radii thus provides a way of evaluating the
stationarity of the flow (see Fig.~\ref{fig:gamma15}). For ${\cal
A}=v_x$, $\Phi_{\cal A}$ represents the momentum accreted in the
$x$-direction. In that case, $\Phi_{\cal A}$ does depend on $R$, since
the velocity $v_x$ of a fluid particle changes due to the
gravitational force of the accreting star. However, the total force on
the star, which is the sum of the accreted momentum and the
gravitational force exerted by the gas on the star, should again be
independent of $R$. Finally, for ${\cal A}=xv_y-yv_x$, $\Phi_{\cal A}$
is the accreted spin momentum.

\subsection{Test simulations}
We tested our code and the prescription for calculating fluxes by
simulating Bondi accretion, i.e. the spherically symmetric accretion
of polytropic gas onto an accretor of constant mass $M$
(Sect.~\ref{Sect:Intro}) at rest.  Given the sound speed $c_\infty$
and density $\rho_\infty$ of the gas at infinity, this problem has a
family of solutions which differ by the effective accretion rate. The
so-called transonic branch corresponds to the situation where the flow
changes from subsonic at large distances to supersonic close to the
accreting object. This solution has the largest accretion rate
(e.g. Theuns \& David 1992). A flow pattern corresponding to this
branch and matching the analytical solution is first set up. The
parameters $M$, $c_\infty$ and $\rho_\infty$ are then scaled in order
to get flows with the same average resolution $\overline{h}$ within
the sphere of radius $r_h$, but with the sonic point falling either
outside the accreting sphere's surface (the `supersonic' model) or
inside the accreting sphere (the `subsonic' model). Note that the
{\it physical model} is identical in both cases, the only difference 
being in the place where mass is removed from the flow: 
in the subsonic model, potentially
all mass is removed from the flow before it has managed to become
supersonic.

\begin{figure*}
\setlength{\unitlength}{1cm}
\centering
\begin{picture}(10,8)
\put(0.0,0.0){\includegraphics{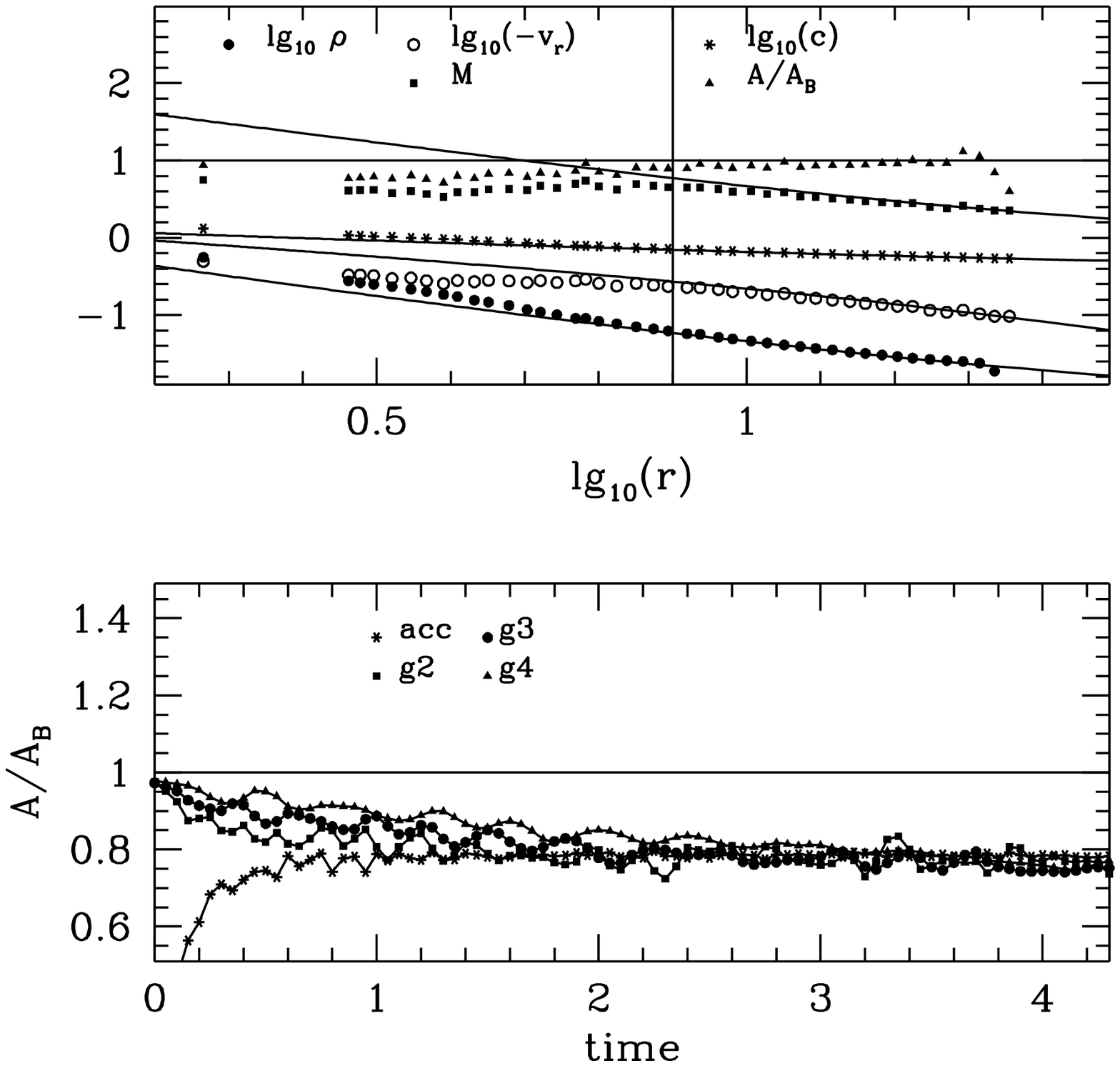}}
\end{picture}
\caption[]{Comparison of SPH test simulation (with $\gamma = 1.5$)
against analytic solution for the transonic branch of Bondi
accretion. The corresponding Bondi radius, as defined in
Eq.(\ref{eq:B}), is 400 in our units. Upper panel: SPH profiles (in
the stationary regime) of density $\rho$, infall velocity $v_r$, sound
speed $c$, mass flux $A=-4\pi r^2\rho v_r$ (normalised by the
analytical value $A_B$) and Mach number $\cal M$ compared with the
analytic solution (solid lines). For the sake of clarity, density,
velocity and sound speed profiles have been offset by -1, -0.7 and
-0.4, respectively. Particles to the left of the vertical line are
used to compute the average resolution $\overline{h}$ around the
accreting object (i.e. $\log_{10}(r_h) = 0.9$).\\ Lower panel:
accretion rates (normalised to the analytical value) determined from
Eq.~(\ref{Eq:dmA}) (curve labelled `acc') compared with rates
determined from Eq.~(\ref{Eq:rhoAv}) for three values of the radius
$R$ ($R=2\overline{h}$, labelled `g2'; $R=3\overline{h}$, labelled
`g3'; $R=4\overline{h}$, labelled `g4', where $\overline{h}\approx
0.8$), as a function of time. For times $\ge2$, a steady-state is
reached.
\label{fig:bondisub}
}
\end{figure*}

The upper panel of Fig.~{\ref{fig:bondisub}} compares the simulation
results for the `subsonic' case with the analytical solution. The
agreement is satisfactory outside the accreting sphere (only there is
the comparison meaningful, obviously).  The deduced accretion rates in
the stationary regime are 20\% low with respect to the analytic value
(lower panel of Fig.~\ref{fig:bondisub}), which is sufficient for our
purpose. The agreement is better in case the sonic point falls outside
the accreting sphere (the \lq supersonic\rq~ model).  Note that the
difference between analytic solution and simulation is likely to be
due, not only to the accuracy of the flux calculation itself, but also
to how well the initial profile matches the analytic profile. The
resolution in these test cases, in terms of the number of particles
inside the region in which particles are being accreted, is comparable
to the resolution obtained in the standard wind accretion case, to be
discussed in Sect.~\ref{Sect:conditions}.

\subsection{Numerical simulations}
\label{Sect:conditions}

All the simulations considered in this paper are for a binary system
consisting of a $3~\Msun$ AGB star and a 1.5~$\Msun$ companion in a
circular orbit around the AGB star with a semi-major axis of 3~AU
(corresponding to a period of 895~d, and an orbital velocity of
36~\kms). The AGB star is losing mass at a steady rate of $\dot{M}_1 =
10\sp{-6}~\Msun$~y$\sp{-1}$ with a (constant) wind velocity of 15~\kms
(here, and in what follows, subscripts 1 will refer to the mass-losing
star). The sound speed at the base of the flow is 7.6~\kms
(corresponding to a temperature of about 4050/$\gamma$~K at the
surface of the AGB star; note that this sound speed is larger than the
value used in Paper~I). Three models are considered which differ only
in the assumed value of $\gamma$ (1.5, 1.1 and 1). Unless stated
otherwise, we use units in which period $P$, semi-major axis $A$ and
gravitational constant are all unity.  In these units, the parameter
$r_h$ (Sect.~\ref{Sect:numerical}) is taken equal to 0.3.

\begin{table*}
 \begin{minipage}{120mm} 
   \caption{Basic data and accretion rates for the different cases considered
     \label{table:runs}}
     \begin{tabular} {@{} rrrrrrrrrr} 
N    & $\gamma$ & $2 \overline{h}/A$ & $2\overline{h}$    & $t/P$   & $\dot{M}_2/\dot{M}_1$ & 
$H_y$ & $G_y$ & $F_y$ & $\dot L_z/V_c A\dot{M}_2$ \\
     &          &  & $\Rsun$               &    &  \%      &$\dot{M}_2 V_c$&
$\dot{M}_2 V_c$&$\dot{M}_2 V_c$&  \\
\hline
100k &   1.5    &   0.11 & 71                  & 24  &   2      & -1.5    
     & 1.55     &   0.05 &     0.06     \\
100k &   1.1    &   0.10 & 64                  & 17  &   8      &  --   
     & --       & --    & 0.06 \\
70k  &   1      &   0.10 & 64                  & 22  &   8      &  --   
     & --       & --    & 0.06 \\
  \end{tabular}
 \end{minipage}
\end{table*}

Table~{\ref{table:runs}} provides the conditions for the different
runs performed and the corresponding accretion rates. The successive
columns give the total number of particles $N$, the polytropic index
$\gamma$, the average resolution $2\overline{h}$ inside the sphere of
radius $r_h$ around the accreting star [as is common practice, we use
$2\overline h$ to characterise the resolution because the SPH kernel
$W(r/h)$ becomes zero for $r>2h$, see Paper~I], expressed relative to
the orbital separation and in solar radii, the number of orbital
periods during which the simulation has been carried out ($t/P$), the
fraction of the mass lost by the AGB star that is accreted by the
companion ($\dot{M}_2/\dot{M}_1$), the hydrodynamic, gravitational and
total drags ($H_y$, $G_y$ and $F_y$, respectively, expressed in terms
of the accreted momentum $\dot{M}_2 V_c$), and finally the
dimensionless spin accretion rate $\dot L_z/V_c A \dot{M}_2$. The
various accretion rates will be described in more details in
Sect.~{\ref{Sect:Results}}. A dash in Table~{\ref{table:runs}} means
that the value has not been computed.

\begin{figure*}
\setlength{\unitlength}{1cm}
\begin{picture}(9,9)
\centering
\put(0.0,0.0){\includegraphics{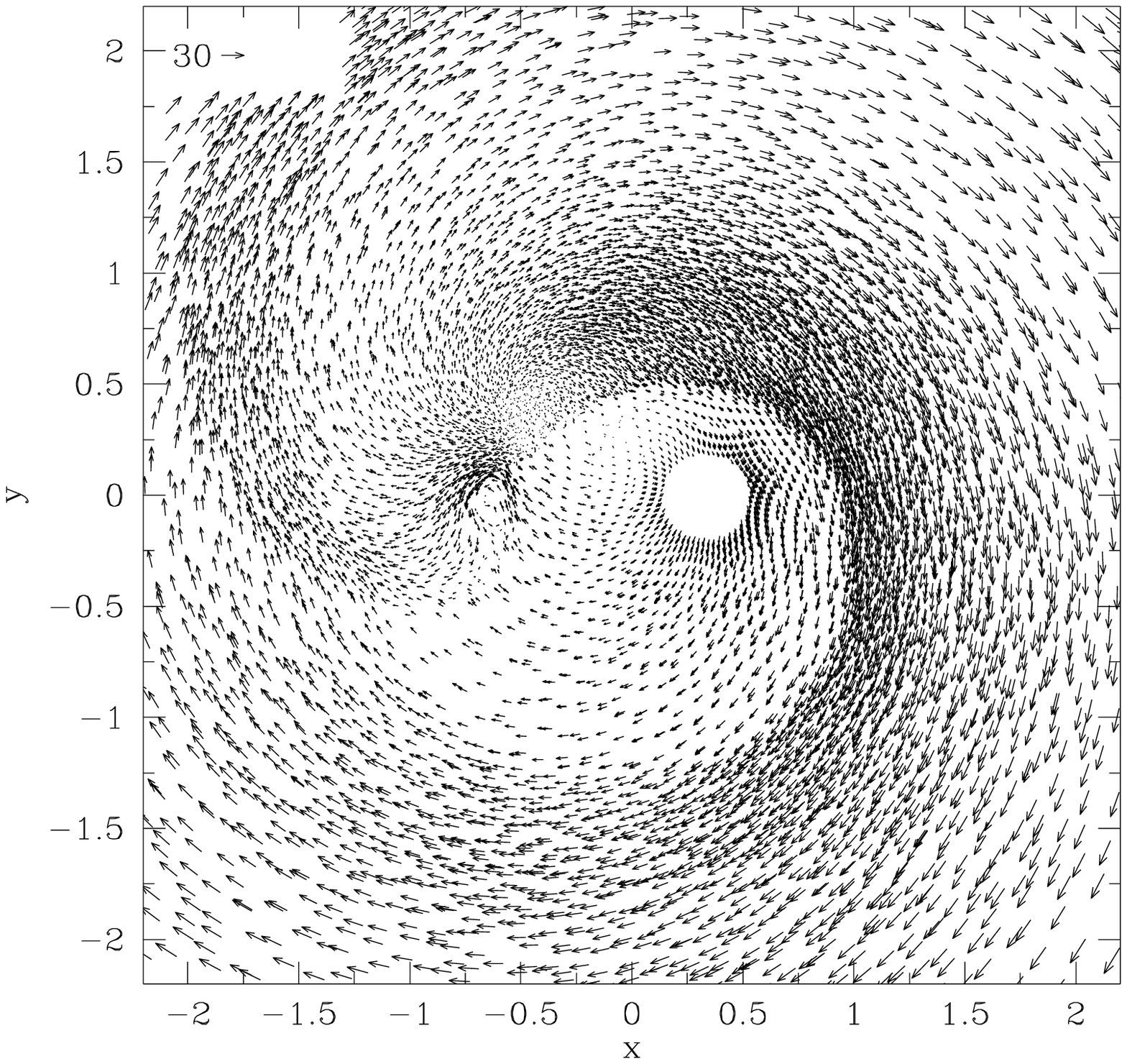}}
\put(0.0,0.0){\includegraphics{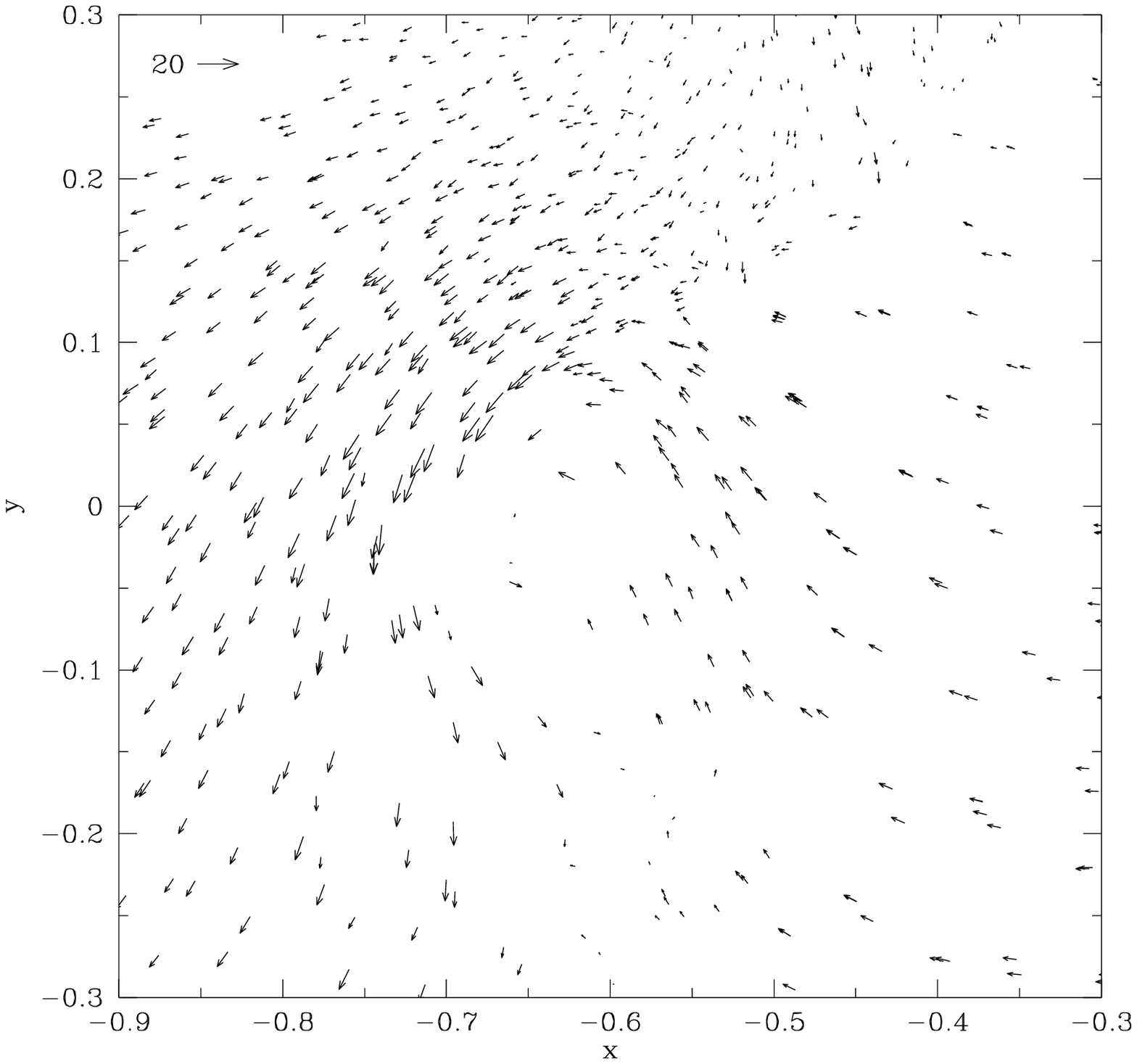}}
\end{picture}
\caption[]{
\label{Fig:flow}
Left panel: General structure of the velocity field in the orbital
plane for the adiabatic case $\gamma = 1.5$ with 100k particles. Only
particles lying in a $z$-slice [-0.1,0.1] enclosing the orbital plane
are represented.  The stationary frame in which the mass-losing star
is at ($x = 0.33, y = 0, z = 0$) and the accreting star at ($x =-0.66,
y = 0, z = 0$) has been used.\\ Right panel: Same as left panel,
zoomed in on the accreting star. The spin momentum carried by the wind
close to the accretor is apparent.}
\end{figure*}

Each simulation typically requires about two weeks of CPU time on a
HP-730 workstation. In terms of the accretion radius $R_A$
[Eq.(\ref{eq:RA})], the resolution achieved around the accreting star
amounts to $2\bar{h}/R_A \sim 0.2$, since $R_A = 380\;\Rsun$.  Since
this resolution is much higher than in Paper~I (100k particles instead
of 40k particles for the case $\gamma = 1.5$), Fig.~\ref{Fig:flow}
displays a general view of the flow pattern, as well as a zoom on the
region close to the accreting star. Despite a larger sound speed at
the surface of the AGB star, the general structure of the flow is
similar to that described in Paper~I. It is worth noting in relation
with the discussion of Sect.~\ref{Sect:barium_mass} that the
Roche-lobe geometry is not at all apparent in the flow structure,
because the wind acceleration mechanism has the net effect of
suppressing the gravitational potential of the AGB star (see
Sect.~\ref{Sect:numerical}).

\section{Results} 
\subsection{SPH accretion rates}
\label{Sect:Results}

The accretion rates of mass, momentum and spin for the three values of
the polytropic index ($\gamma = 1, 1.1$ and 1.5) considered in this
paper are given in Table~{\ref{table:runs}}. They will be discussed in
some detail here for the $\gamma=1.5$ (adiabatic) case only, since the
same remarks apply to the other cases as well.

\begin{figure*}
\setlength{\unitlength}{1cm}
\centering
\begin{picture}(12,8.49)
\put(0.0,0.0){\includegraphics{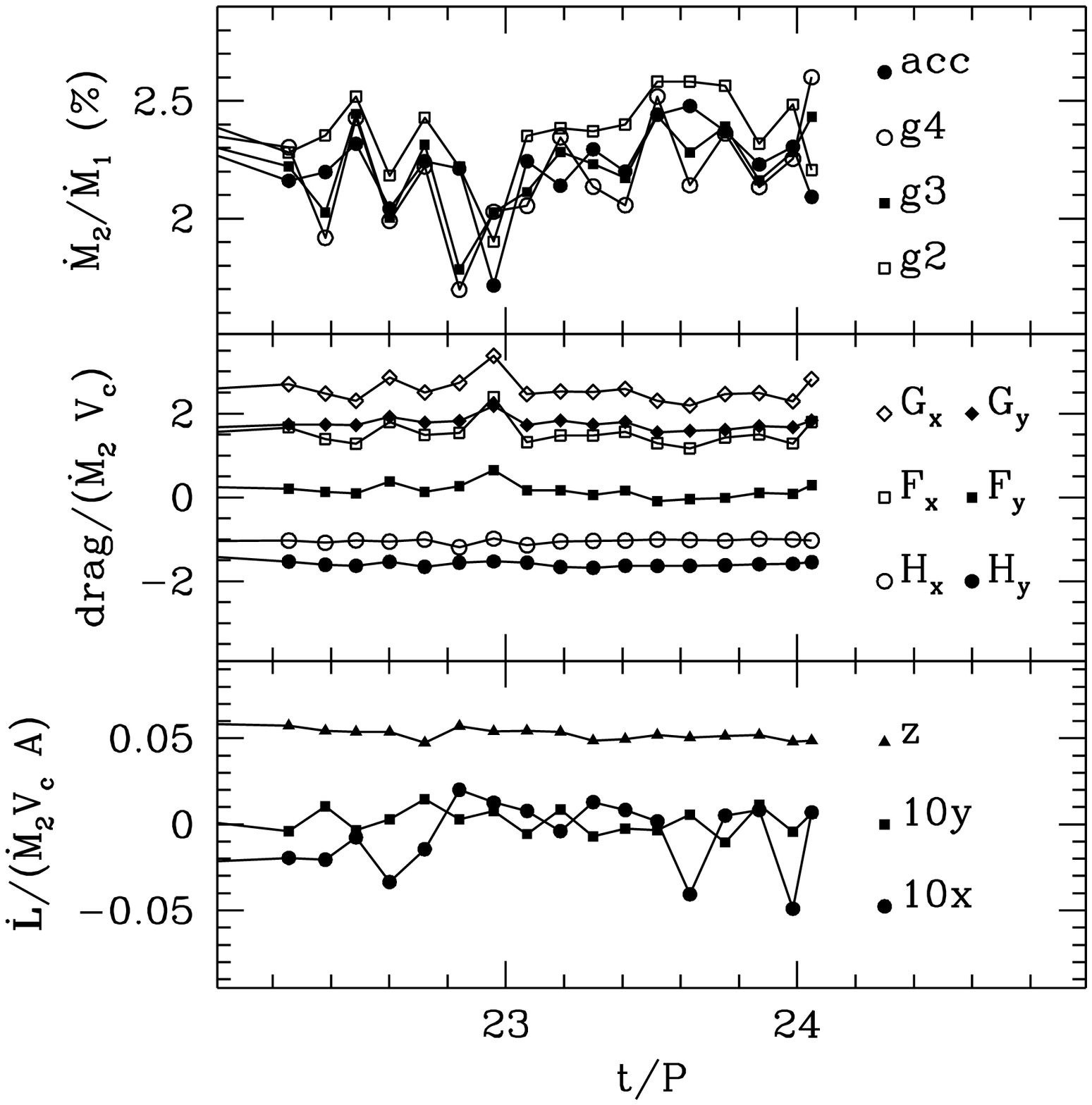}}
\end{picture}
\caption[]{ Accretion rates as a function of time for the $\gamma=1.5$
case.\\ Top panel: accretion rates of mass ($\dot{M}_2/\dot{M}_1$, in
\%).  The different symbols represent estimates based on
Eq.~(\ref{Eq:dmA}) (filled dots, labeled `acc') or on
Eq.~(\ref{Eq:rhoAv}) with three different radii $R$ (0.11, 0.16 and
0.22, labeled `g2', `g3' and `g4', respectively).\\ Middle panel:
accretion rates of linear momentum in the radial (subscript $x$) and
tangential ($y$) directions, in units of $\dot M_2 V_c$: $H$ refers to
the hydrodynamic drag, $G$ to the gravitational drag and $F$ to the
sum of those (see text for details).\\ Bottom panel: rate of accretion
of dimensionless spin momentum, $\dot L/\dot M_2 V_c A$, in the radial
$x$, tangential $y$ and $z$ directions. Spin momenta along $x$ and $y$
are multiplied by ten.}
\label{fig:gamma15}
\end{figure*}

Figure~{\ref{fig:gamma15}} shows the accretion rates as a function of
time. The top panel indicates that the two methods of determining the
fluxes, namely direct summation over SPH particles (Eq.~\ref{Eq:dmA})
or estimates of SPH fluxes (Eq.~\ref{Eq:rhoAv}), give very similar
results: about 2 to 2.5\% of the mass lost by the AGB star is accreted
by the companion (but see below). Incidentally, the fact that the mass
fluxes through spheres of different radii are almost equal indicates
that the flow pattern is relatively stationary.

The middle panel gives the forces exerted by the gas on the accreting
star, due to the accretion of linear momentum (hydrodynamic drag,
labeled $H$) and due to the gravitational pull of the gas lagging
behind the accretor (labeled $G$). More precisely, the accretion of
linear momentum has been computed from
\begin{equation} 
{\bf H} = {\PP - \dot M_2 \vv_2},
\end{equation}
where $\PP \equiv \sum_w \vv_w \del m_w /\del t$ denotes the momentum
in the wind, accreted by the star. The index $w$ indicates that the
sum extends over SPH particles. The inertial forces in this figure are
projected on a set of axes, with the $x$ axis along the line joining
the two stars, the mass-losing star lying on the $+x$ axis with its
orbital motion pointing along $+y$, and finally the $z$ axis being
perpendicular to the orbital plane.  The gravitational drag $G$, on
the other hand, is computed by summing over SPH particles in the wind
as:
\begin{equation}
{\bf G} = \sum_w \F_{grav,w}.
\end{equation}
The net force acting on the star is labeled $F\equiv G + H$.

The hydrodynamic drag is negative, for both radial and tangential
directions. This means that the accreting star is propelled away from
its companion and is accelerated along its orbit due to the gas
raining back onto it from $y>0$ (i.e., a thrust rather than a drag),
as apparent from Fig.~\ref{Fig:flow}. However, the gravitational
forces have exactly the opposite tendency and the net effect is an
inward pull and a very slight slowing down of the star along its
orbit.

The bottom panel shows the accretion of the dimensionless spin angular
momentum, again obtained by summing over SPH particles:
\begin{equation}
{{\bf\dot{L}}\over A\,V_c\,\dot M_2} = \sum_w ({\bf r}_w - 
{\bf r}_2) \times (\vv_w - \vv_2) 
{1\over A\,V_c\,\dot M_2}\,{\del m_w\over \del t}
\end{equation}
The radial ($x$) and tangential ($y$) components fluctuate around
zero, as expected from the symmetry of the flow along the $z$
direction (perpendicular to the orbital plane).  There is, however, a
net accretion of spin momentum in the $z$ direction, in such a way
that the accreting star tends to be spun up if it is in synchronous
rotation to begin with. This can be clearly seen on
Fig.~\ref{Fig:flow} from the direction of rotation of the disc-like
structure around the accreting star. The accretion rates of
dimensionless spin momentum are remarkably similar amongst the three
models, notwithstanding the fact that a much better-developed
accretion disc forms in the isothermal case (compare Figs.~1 and 8 of
Paper~I).

\begin{figure*}
\setlength{\unitlength}{1cm}
\centering
\begin{picture}(12,8.49)
\put(0.0,0.0){\includegraphics{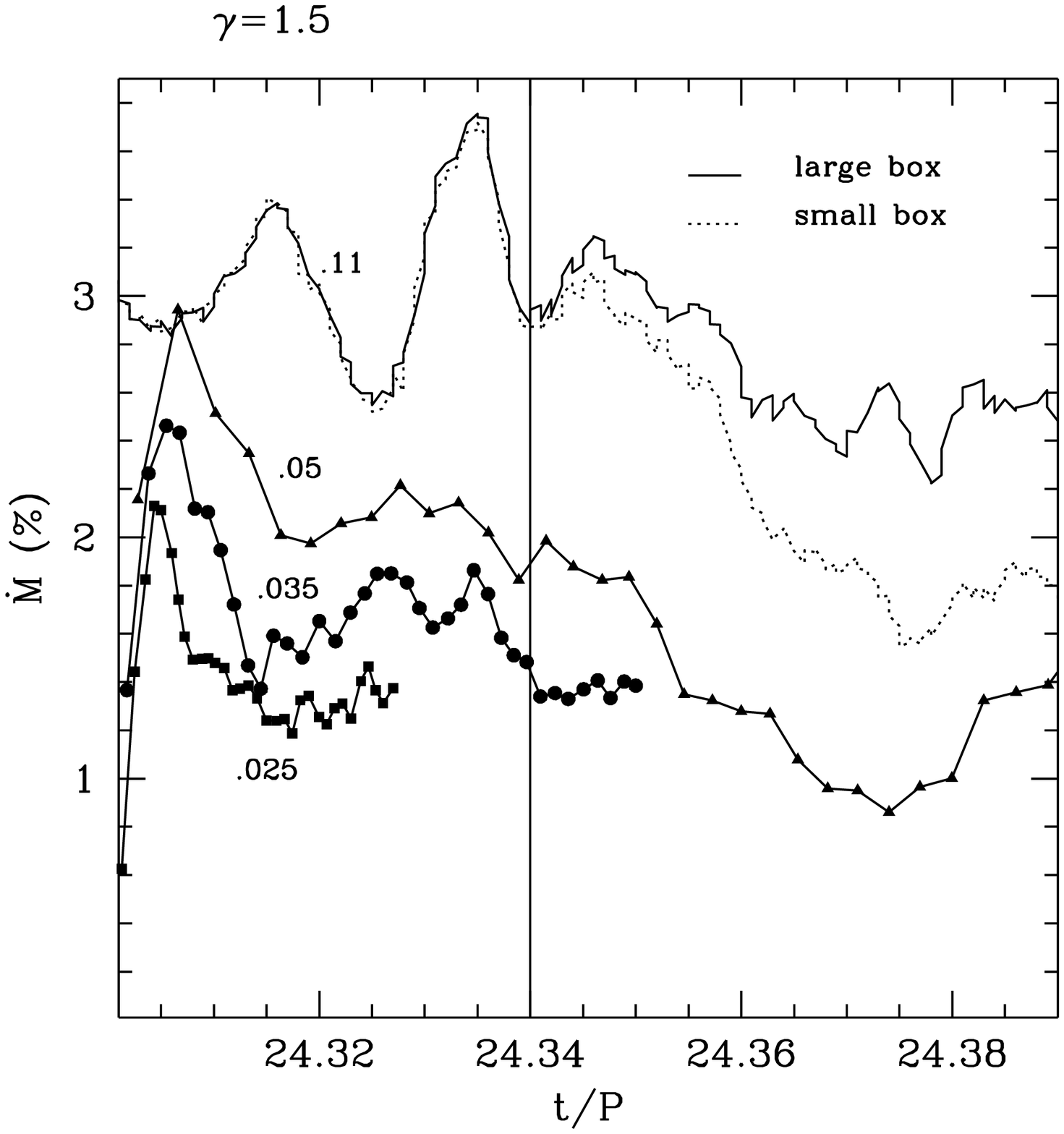}}
\end{picture}
\caption{ Accretion rates of mass ($\dot{M}_2/\dot{M}_1$, in \%) as a
function of time (in units of the orbital period $P$) for the
$\gamma=1.5$ model with different resolutions $2\bar h$. The solid curve labeled
\lq large box\rq~ refers to the standard model. The dotted line
labeled \lq small box\rq~ corresponds to a simulation where only those
particles close to the accretor are retained. Up to $t/P \approx
24.34$, these two rates are identical. The other curves (triangles,
dots and squares) refer to simulations in a \lq small box\rq~ as
above, but with an initial number of particles increased as indicated
in the text. The different curves are labeled with their effective
resolutions.}
\label{fig:highres}
\end{figure*}

\begin{figure}
\setlength{\unitlength}{1cm}
\centering
\begin{picture}(8,5)
\put(0.0,0.0){\includegraphics{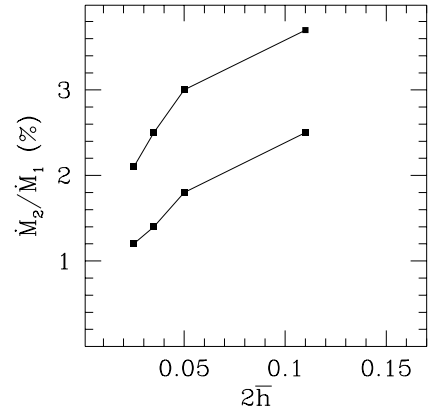}}
\end{picture}
\caption{Accretion rates of mass ($\dot{M}_2/\dot{M}_1$, in \%) as a
function of resolution, $2\bar h$, for the $\gamma=1.5$ model. The
upper and lower curves correspond to the maximum and minimum accretion
rates for these resolutions, read from Fig.~\protect\ref{fig:highres}.}
\label{fig:minmax}
\end{figure}

We tried to improve our estimate of $\dot {M}_2/\dot {M}_1$ by
increasing the resolution drastically in a small region around the
accreting star. However, conclusions drawn from simulations involving
such a restricted region will be meaningless once boundary effects
propagating inwards become important. In order to illustrate that
effect, a simulation has been performed with most of the particles
removed, except for those close to the accreting star. As illustrated
by Fig.~\ref{fig:highres}, the accretion rate does not change when
measured over a sufficiently short interval of time, but starts
deviating when boundary effects propagate into the accretion sphere
(round $t/P=24.34$). In Fig.~\ref{fig:highres}, the line labeled \lq
large box\rq\ corresponds to the accretion rate for the standard
model\footnote{The reason why the standard case yields accretion rates
of the order of 2.5--3.5\% in Fig.~\ref{fig:highres} as compared to
2.0--2.5\% in Fig.~\ref{fig:gamma15} is because the latter correspond
to time averages.}, whereas the dotted line, labeled `small box',
includes particles close to the accreting star only (in fact, the
small box {\it with the same} resolution as the large box, contains
only 4\% of the number of particles of the large box). Obviously, for
$t/P \le 24.34$, the accretion rates remain identical yet start to
diverge afterwards. Therefore, the high-resolution simulations
described below, and involving the small box, are only meaningful for
$t/P \le 24.34$. In order to artificially increase the resolution, we
proceed as follows. At time $t/P = 24.30$, each SPH particle is
replaced by a swarm of $n$ particles, distributed around their mother
according to the SPH density kernel. This increases the resolution by
a factor $n^{1/3}$. The different curves in Fig.~\ref{fig:highres}
correspond to simulations with $n=1$ (dotted), $n=10$ (triangles),
$n=25$ (dots) and $n=90$ (squares), which translates into a number of
particles in the small box around the accreting star of 4$k$, 40$k$,
100$k$ and 360$k$, respectively. Drawing conclusions from these
numerical experiments is hampered by the fact that the accretion rate
varies quite considerably over time intervals of the order of 0.02$P$,
as is clear from the $n = 1$ case (solid line). These variations on a
time scale much longer than the time step $\delta t\le 5\,\times
10^{-4}P$ simply reflect the discrete episodes of particle ejections
by the mass-losing star (see Sect~2.3 of Paper~I). In addition, once
SPH particles are cloned, as described above, the system needs some
time to relax to a new steady state.  The following conclusions can
nevertheless be drawn from these simulations: 1) the accretion rate
{\it decreases} with increasing resolution, and 2) the mass accretion
rate is not likely to be much less than about 1\%, i.e. about 1/3
lower than deduced from the standard model. This value follows
assuming that the almost linear trend observed in
Fig.~\ref{fig:minmax}, which displays numerical resolution versus mass
accretion rate, can be extrapolated to the size of a main sequence
accretor (i.e., $2\bar h\approx 1.5\,\times 10^{-3}$ which corresponds
to 1$\Rsun$). In what follows we will use an estimated mass accretion
rate of 1\% for the adiabatic model, but we stress that
an accurate estimate of the real accretion rate requires much higher
resolution simulations than the ones presented here.

\subsection{Comparison with the Bondi-Hoyle mass accretion rate}
\label{Sect:comparison}
In Sect.~\ref{Sect:Results}, a mass accretion rate $\beta = -
dM_2/dM_1$ of the order 1\% is derived from our SPH simulations in the
$\gamma=1.5$ case. That result will now be compared to the accretion
rate derived by the Bondi \& Hoyle (1944) formula.  In the framework
of a binary system, the gas velocity $v_\infty$ appearing in
Eq.~(\ref{Eq:BH}) may be replaced by the relative velocity between the
wind and the accreting star, whereas the gas density $\rho_\infty$ may
be replaced by the wind density $\rho$ at the location of the
accreting star. The latter may be deduced from the mass conservation
equation $\dot{M}_1 = -4\pi r^2 v_{\rm w} \rho$. Substituting these
expressions into Eq.~(\ref{Eq:BH}) yields
\begin{equation}
\label{Eq:BHB}
\beta =  \frac{\alpha}{A\sp2}
\left(\frac{GM\sb2}{v\sb{\rm w}\sp2}\right)\sp2 
\frac{1}{[1+(v\sb{\rm orb}/v\sb{\rm w})\sp2 + (c/v_{\rm w})\sp2]\sp{3/2}}, 
\end{equation}
or using Kepler's third law:
\begin{equation}
\beta = \alpha\, \mu ^2 \, {k^4 \over {[1+k^2+(c/v_{\rm w})\sp2]^{1.5}}},
\label{eq:BH2}
\end{equation}
where $\alpha$ is a parameter of order unity not fixed by the theory,
$\mu \equiv M_2 / (M_1 + M_2)$, $k \equiv v_{\rm orb}/v_{\rm w}$, and
$v_{\rm orb}$ is the orbital velocity.

For the conditions given in Sect.~\ref{Sect:conditions},
Eq.~(\ref{eq:BH2}) yields $\beta \simeq 10$ to 20\% for $k = 2.4$, taking
$\alpha$ between 0.5 and 1, as suggested by the numerical simulations
of Ruffert \& Arnett (1994). The accretion rate deduced from the SPH
simulation in the $\gamma = 1.5$ case is thus about ten times smaller than
that predicted by the Bondi \& Hoyle formula. The reason for this
apparent discrepancy is that Eq.~(\ref{eq:BH2}) is only applicable to
situations where $k << 1$, since it was originally derived by
considering a single star moving through a gas cloud. In the case of a
fast wind ($k << 1$), the concept of accretion column as introduced
originally by Bondi \& Hoyle to describe the accretion process, does
hold with the accretion column being tilted with respect to the radius
vector. When $k$ is of the order of unity, however, as is the case
here, the binary motion strongly disturbs the shape of the accretion
column, as is obvious from Fig.~\ref{Fig:flow}. The fact that the
above formula becomes meaningless for large values of $k$ is further
illustrated by the fact that $\beta$ increases like $k$ when $k>>1$,
though values of $\beta$ larger than unity have of course no physical
meaning. The fact that the Bondi-Hoyle formula Eq.~(\ref{eq:BH2}) was
not expected to hold in the case of accretion of a slow wind was the prime
motivation to perform the simulations described here.

Finally, note that the term $(c/v_{\rm w})\sp2$ in Eq.~(\ref{eq:BH2})
is usually negligible (in the case of adiabatic wind expansion, $c$
decreases as $r^{-(\gamma - 1)})$, and has therefore not been
considered in the above discussion.
 
\section{Consequences of wind accretion}
\label{Sect:consequences}

Wind accretion can alter the evolution of a binary in several respects:
\begin{itemize}
\item the chemical composition of the envelope of the accreting star may
be modified if the accreted matter has a different composition, and if
enough such mass is accreted;
\item the orbital elements may change as a result of the mass loss
and mass transfer;
\item the accreting star may be spun up if the accreted matter
possesses significant angular momentum.
\end{itemize}

Each of these effects is now considered in turn on general grounds.
Application to barium stars will be considered in
Sect.~\ref{Sect:barium}.

\subsection{Envelope pollution}

The pollution of the envelope of the accreting star depends on (i) the
amount of mass $\Delta M_2$ accreted from the wind, (ii) the dilution
factor of the accreted matter in the envelope, and (iii) the extent to
which the chemical composition of the accreted matter differs from
that of the envelope.  We will assume that the accreted matter has
been fully mixed in the envelope. In the case that the accreting star
is on the giant branch, such a mixing naturally results from the
convective nature of the envelope. In the other case where the
accreting star is still on the main sequence, some mixing will occur
if the accreted matter has a mean molecular weight larger than that of
the underlying layers (Proffitt 1989; Proffitt \& Michaud 1989). Such
an inversion of mean molecular weight, triggering a turbulent mixing
until the inversion is suppressed, is expected if carbon-rich matter
from the wind of an AGB star falls on top of the surface layers of the
main sequence star.

The overabundance $f_i$ of element $i$ in the envelope of the
accreting star (i.e. the ratio between the abundance after completion of
the accretion and mixing processes, and the abundance in the
primordial envelope) is related to its overabundance $g_i$ in the wind
(i.e. in the AGB atmosphere) through the relation
\begin{equation}
\label{Eq:over}
f_i = \frac{         g_i  \Delta M_2 + M_{2,0} - M_{\rm 2,core} }
           {\phantom{g_i} \Delta M_2 + M_{2,0} - M_{\rm 2,core} } 
   \equiv g_i {\cal F} + (1 -  {\cal F}),
\end{equation}
where ${\cal F}$ is the dilution factor of the accreted matter $\Delta
M_2$ in the envelope of mass $(M_2 - M_{\rm 2,core})$. Here,
$M_{2,0}$ and $M_{\rm 2,core}$ denote the total initial and core mass,
respectively, of the accreting star. In deriving the above relation,
it has been assumed that gas and grains are accreted with the same
efficiency. The chemical fractionation that occurs between the gas and
grain phases in the AGB wind (due to the different refractory
properties of the different chemical elements) is then erased once
they mix again in the companion's envelope. This assumption is
actually not always satisfied: in some observed post-AGB star binaries
only gas and no grains have been accreted (e.g. Waters et al. 1992).

\subsection{Variation of orbital parameters}
\label{Sect:orbital}

The orbital parameters of the binary will change as a consequence of
the mass lost from the system, the mass transferred from one star to
the other, the gravitational force of the wind on both stars and
finally, the change in momentum of the accretor, due to the accretion
of momentum from the wind. In addition, the mass loss process itself
may be asymmetric, providing an extra source of momentum 
changing the orbital parameters, but that possibility will not be 
considered here.

Huang (1956) derives expressions for such changes in an adiabatic
approximation, i.e., when the change per period is small. 
However, Huang's parametrization does not allow to compare the
respective effects of drag and mass accretion on the
variations of the orbital elements, though the former effect is of importance
in the situation modelled in this paper (as can be judged from
Fig.~\ref{fig:gamma15}). Therefore, Eqs.~(\ref{eq:dAdt} --
\ref{eq:dedt}) describing the variation 
of the orbital elements are parametrized in term of
$F_y$, the tangential component (i.e. perpendicular to the
radius-vector, and positive along the orbital motion of the
mass-losing star~1) of the total force, 
including gravity and accretion of linear momentum. With the notations of
Sect.~\ref{Sect:Results}, $F_y = G_{2,y} - (M_2/M_1) G_{1,y} + H_y$,
as displayed in Fig.~\ref{fig:gamma15} (but note that we neglected the
gravitational force $G_{1,y}$ exerted by the gas on the mass-losing
star in that figure). That quantity, as well as the mass accretion
rate, are assumed to remain constant during one orbital cycle:
this is probably a good approximation
as long as $e\ll 1$ ($e$ is the eccentricity of the orbit).
Following this line of argument (see the Appendix for more details), 
we find, neglecting all terms of
order $e^2$ or higher:
\begin{eqnarray}
\frac{\dot A}{A} & = & -\phantom{2}{\dot M_1+\dot M_2\over M_1+M_2}
   - 2 {F_y\over M_2 V_c}
\label{eq:dAdt}\\
{\dot P\over P} & = & -2 {\dot M_1+\dot M_2\over M_1+M_2} 
     -3 {F_y\over M_2 V_c}
\label{eq:dPdt}\\
{\dot e\over e} & = &
-{1\over 2} {\dot M_2\over M_2} +{3\over 2} {F_y\over M_2 V_c}.
\label{eq:dedt}
\end{eqnarray}
Equations (\ref{eq:dAdt}) and (\ref{eq:dPdt})
illustrate basically the same physical effects: mass loss from the
system ($\dot M_1+\dot M_2<0$) increases $A$ and $P$ (first term), and
slowing down of the accretor or speeding up of the mass-loser due to
the exerted forces ($F_y>0$) has the opposite effect (last term). The
behaviour of the eccentricity can be written slightly more revealing
in the case of equal masses:
\begin{eqnarray}
{\dot e\over e } & = &{3\over 2} ({1\over 6}+{v_{w,y}\over
V_c}) {\dot M_2\over M_2} + {3\over 2} {G_{2,y}-G_{1,y}\over M_2 V_c},
\nonumber\\
&& {({\rm for\,}\;M_1=M_2)}
\label{eq:dedt2}
\end{eqnarray}
where $v_{w,y}$ is the tangential wind velocity, at the point it is
accreted. (We have replaced $v_{2,y}$ by $-V_c/2$ in doing this
transformation, which is a valid approximation since the eccentricity
is assumed to be small.) Obviously, if the orbit is circular to begin
with, it stays circular. Next, in the absence of accretion ($\dot
M_2=0$), the eccentricity will ${\it increase}$, since most of the
wind material is lagging behind star 2 ($G_{2,y}>G_{1,y}>0$), as can
be seen on Fig.~1b of Paper~I. Hence the second term is
positive. Finally it will depend on the value of the accretion rate
$\dot M_2$ and the wind velocity at the point it is accreted, whether
$e$ will increase (second term dominates) or decrease (first term
dominates).

Applying Eq.~(\ref{eq:dedt}) to our case, we find that the
eccentricity {\it decreases} for our set of parameters (see
Fig.~\ref{fig:gamma15}), its relative variation being of the order of
$-\del M_2/M_2$, i.e. a few percent only. Including the
gravitational force due to the gas on the mass-losing star strengthens
this conclusion.

\subsection{Influence on spin}
\label{sect:spinup}
The companion star may accrete spin angular momentum from the wind,
which may alter its rotational velocity. Packett (1981) showed that a
star needs to accrete only a few percent of its own mass from a
keplerian accretion disc to be spun up to the equatorial centrifugal
limit, essentially because stellar moments of inertia are generally
much smaller than $MR^2$. Less drastic effects than centrifugal
breakup could already be detectable on a statistical basis by
comparing the rotational velocities of post-mass transfer stars with
those of single stars of the same age, thus suggesting a further
observable diagnostic of mass transfer.

The initial spin angular momentum $S_{2,0}$ of the star is defined in
terms of its moment of inertia $I_{2,0}$ through
\begin{eqnarray}
S_{2,0} & = & I_{2,0}\ \omega_{2,0}\label{eq:S}\\ 
I_2 & = & M_2 r_g^2 R^2_2\label{eq:I_2},
\end{eqnarray}
where $I_{2,0}$ is the initial value of $I_2$. The dimensionless
quantity $r_g$ is the gyration radius. The initial spin velocity of
the star is related to its initial rotational velocity through
$\omega_{2,0} = v_{\rm rot,0}/R_2$, with $R_2$ the stellar radius. The
accretion of angular momentum $\Delta L$ will change the angular
velocity from $\omega_{2,0}$ to $\omega_{2} = (S_{2,0} + \Delta L)
(M_{2,0} + \Delta M_2)^{-1} r_g^{-2} R^{-2}_2$, assuming that 
$r_g$ and $R_2$ did not change much as a result of the accretion process. 
This value should not exceed the critical centrifugal
angular velocity $\omega_{\rm cr} = [G (M_2+\Delta M_2)/R_2^3]^{1/2}$,
with $G$ the gravitational constant. In terms of
the accreted spin momentum $\Delta L \Delta M_2^{-1} A^{-1} V_c^{-1}$
as defined in Sect.~\ref{Sect:Results}, the rotation velocity $v_{\rm
rot}$ after spin accretion writes
\begin{equation}
\label{Eq:vrot}
v_{\rm rot} = (v_{\rm rot,0} + \frac{\Delta L}{\Delta M_2 A V_c}
\frac{A}{R_2} \frac{1}{r^2_g} \frac{\Delta M_2}{M_{2,0}}V_c)/(1 +
\frac{\Delta M_2}{M_{2,0}}).
\end{equation}
The above equation implicitly assumes that the accreted spin has been 
redistributed over the entire envelope, which is equivalent to assuming that
the accreted matter has been fully mixed in the convective envelope. 

\section{Barium stars}
\label{Sect:barium}

Though the simulations described in this paper could be applied to a
variety of binary systems, they were aimed at gaining some insight in
the origin of barium stars. In particular they aim at checking the
validity of the wind accretion model for the formation of barium stars
developed by Boffin \& Jorissen (1988). The impact of our hydrodynamic
results on the validity of that scenario will therefore be briefly
discussed in this section.

In order to evaluate the total variation of the orbital elements due
to the wind accretion process, Eqs.~(\ref{eq:dAdt})-(\ref{eq:dedt})
have to be integrated with $M_1$ as the independent variable until the
mass of the AGB primary has been reduced from its initial value
$M_{1,0}$ to that of its WD descendant. The total amount of mass lost
by the primary during its evolution from AGB to WD, $\Delta M_1$, can
be derived from the initial -- final mass relationship. According to
Weidemann (1984), a star of initial mass $M_{1,0}$ will give rise to a
WD of mass $M_1^{WD}$, such that
\begin{equation}
\label{Eq:WD}
M_1^{WD} = 0.528 - 0.044 M_{1,0} + 0.016 M_{1,0}^2.
\end{equation}
It is assumed that this relation, derived for single WD's, holds true
as well for WD's in {\it detached} binary systems where mass loss
occurred through a wind.

To obtain changes in composition and changes in orbital parameters, we
start with $M_1 = M_{1,0}$ and integrate Eqs.~(\ref{eq:BH2}),
(\ref{eq:dAdt}), (\ref{eq:dPdt}) and (\ref{eq:dedt}) to find $M_2$,
$A$, $P$ and $e$. The integration is stopped when the mass-losing star
reaches its WD mass. The total amount of mass accreted by the barium
star, $\Delta M_2$, can then be used to predict the dilution factor
$\cal{F}$ appearing in Eq.~(\ref{Eq:over}) and, hence, the pollution
of the barium star envelope.

\subsection{Mass accretion rate}
\label{Sect:barium_mass}

The major result of the present paper is that the mass accretion rate
obtained from our SPH simulations is significantly lower than the
commonly used Bondi-Hoyle value. The extent to which such a
reduction affects the ability of the wind accretion scenario to
account for the chemical peculiarities of barium stars is now
discussed.

As a first order approximation we assume that the accretion efficiency
remains the same during the whole AGB mass loss episode, i.e. $\beta =
0.01$ (as discussed in Sect.~\ref{Sect:Results}) throughout. In that case,
of the $\sim 2.45~\Msun$ lost by the $3~\Msun$ AGB primary [see
Eq.~(\ref{Eq:WD})], 0.02~$\Msun$ will be accreted by the
companion. This enriched material will be diluted in a $\sim
1.1~\Msun$ envelope (i.e. adopting 0.4~$\Msun$ for the core mass),
yielding a dilution factor ${\cal F} = 0.02$. Assuming typical
heavy-element overabundances of the order of $g = 100$ in the AGB wind
(e.g. Utsumi 1985), we find a final overabundance $\log_{10} f =
0.5$~dex from Eq.~(\ref{Eq:over}), quite a typical overabundance
factor for barium stars (see e.g. Lambert 1985). In the following,
dilution factors $\cal F$ will always be converted into overabundances
$f$ by means of Eq.~(\ref{Eq:over}) and adopting $g = 100$ and $M_{\rm
2,core} = 0.4\; \Msun$.

The accretion efficiency $\beta$ will probably not remain constant
during the wind accretion process, since it very likely depends upon
the orbital separation (through the ratio $k$ of orbital to wind
velocity) and upon the masses of the two stars, as suggested by
Eq.~(\ref{eq:BH2}) derived in the framework of the Bondi-Hoyle theory.
It is therefore necessary to take into account the variations of the
parameters $k$ and $\mu$ resulting from the dynamics of the mass
transfer process, as outlined in Sect.~\ref{Sect:comparison}.
Unfortunately, the small number of SPH simulations performed prevents
us from checking the functional dependence between $\beta$, $k$ and
$\mu$ suggested by Eq.~(\ref{eq:BH2}). Consequently, we are restricted
to use Eq.~(\ref{eq:BH2}), yet we normalize it in such a way as to
yield our SPH result $\beta = 0.01$ when $v\sb{\rm w} = 15$~\kms, $A =
3$~AU, $M_1 = 3\; \Msun$ and $M_2 = 1.5\; \Msun$.

Starting with the initial parameters as listed above, the average
$\beta$ over the whole mass transfer process is found to be 0.011,
i.e., slightly larger than in the first order approximation 
where orbital parameters are kept constant
(which is equivalent to fixing $\beta$ at its initial value
of 0.010). In other
words, the accretion rate {\it increases} as the system widens due to
the mass lost from it. This may seem rather counter-intuitive at
first, but is easily derived from Eq.~(\ref{Eq:BHB}) provided that  
$v_{\rm wind} << v_{\rm orb}$: $\beta \propto A^{-2} v_{\rm orb}^{-3}
\propto A^{-1/2} (M_1 + M_2)^{-3/2}$ making use of Kepler's third law,
and finally $\beta \propto A$ since Eq.~(\ref{eq:dAdt}) predicts $A
\propto (M_1 + M_2)^{-1}$ during mass transfer when drag effects are
negligible.   
Note that this behaviour holds as long as
the wind velocity is negligible with respect to the orbital velocity.

For wide binaries, smaller $\beta$ values prevail, since one can now
neglect the orbital velocity with respect to the wind velocity to find
that $\beta$ scales as $1/A^2$ according to Eq.~(\ref{Eq:BHB}). It
is therefore expected that wind accretion will be less efficient in
wider systems. The sensitivity upon $k$ predicted by
Eq.~(\ref{eq:BH2}) appears to be rather small, however, since it
predicts average $\beta$ values of 0.011, 0.009 and 0.007 for initial
orbital separations of 3, 4 and 5~AU, respectively, all other
parameters being as listed above. A substantial pollution of the
companion's envelope will still result from these values of $\beta$,
since the corresponding dilution factors ${\cal F} = 0.023$, 0.018 and
0.014, respectively, yield $\log_{10}f=0.52$, 0.44 and 0.38~dex.

The threshold for producing a barium star in a binary system may be
set to ${\cal F} \ge 0.005$, yielding overabundances $\log_{10} f \ge
0.17$~dex, according to Eq.~(\ref{Eq:over}). Below this value, the
weakly polluted star would no longer be flagged as a barium star. With
our new SPH scaling of Eq.~(\ref{Eq:BHB}), we find that the dilution
factor $\cal F$ drops below 0.005 for systems with final orbital
periods in excess of 90~y (adopting the same parameters as above,
except for the initial separation). The barium star with the longest
known orbital period, $\zeta$~Cygni ($P = 17.8$~y, Griffin 1991),
exhibits an average overabundance of heavy elements amounting to
0.4~dex (\Zacs\ 1994). Such an overabundance can easily be accounted
for with the above prescriptions, which predict ${\cal F} = 0.016$ and
$\log_{10}f = 0.4$~dex as required.

{\it We conclude that the reduction in the wind accretion efficiency
derived from our detailed SPH simulations when compared to the
Bondi-Hoyle prescription does not endanger the validity of the wind
accretion scenario.} In particular we find that wind accretion can
easily account for the level of pollution in barium stars with orbital
periods as long as 90~y. In fact, Boffin \& \Zacs\ (1994) already suggested
that a reduction of the accretion rate by a factor of ten with respect to the
Bondi-Hoyle value improves the agreement of wind accretion predictions
with observed levels of enrichement in barium stars.

However, that scenario faces serious difficulties for systems with
{\it short} orbital periods, albeit on different grounds.  Indeed,
barium systems with orbital periods of a few hundred days could not
stay detached when the primary star evolved up the AGB. Standard
prescriptions for AGB evolution (e.g. Groenewegen \& de Jong 1993)
indicate that the mass-losing star will overflow its Roche lobe at
some point during its AGB evolution in systems with $P \le
3000$~d. Since the star has a deep convective envelope at that stage,
the severe mass loss due to the Roche lobe overflow (RLOF) will cause
its radius to expand. But the Roche lobe tends to shrink as a result
of mass transfer from the more massive to the less massive component,
thus leading to a runaway behaviour of the mass transfer process. A
common envelope surrounding both stars is expected to result from this
process, and the drag exerted by the envelope on the binary motion
will cause a severe loss of orbital energy. Binary systems with very
short orbital periods (a few hours), like e.g. cataclysmic variables,
are formed in such a way (e.g. Meyer \& Meyer-Hofmeister 1979). Barium
systems with periods of a few hundred days must somehow have escaped
this dramatic fate, though these systems are not wide enough to
prevent an episode of RLOF. The problem of the short period barium
systems has been illustrated by the recent analysis of HD~121447
(Jorissen et al. 1995a), the barium star with the shortest orbital
period (185~d). Possible solutions are investigated by Han et
al. (1995).

\subsection{Momentum accretion}
Boffin et al. (1993) compared orbital elements of barium stars against
those of normal G-K giants. They concluded that the ($e, \log P$)
diagram of barium stars could be reproduced by translating to larger
periods the points representing normal giants. An increase of the
orbital period is easily explained by mass loss from the system [see
Eq.~(\ref{eq:dAdt})], but the change in eccentricity depends on both
the accretion rate and the influence of the gravitational and
hydrodynamic forces [Eq.~({\ref{eq:dedt}})]. Unfortunately, the latter
equation is probably only valid for nearly circular orbits (for which
we find a small decrease in eccentricity). Hence, we cannot draw
definite conclusions about d$e$/d$t$ for systems which start with a
finite eccentricity. In addition we caution that it is very unlikely
that the barium stars with the shortest orbital periods result from
the widening of systems with even shorter periods, in view of the
difficulties raised by RLOF (see Sect.~\ref{Sect:barium_mass}).

\subsection{Spin accretion}
The mild barium star HD~165141 possibly bears the signature of spin up
due to angular momentum accretion during the mass transfer process
(see Jorissen et al. 1995b for a detailed discussion of that system).
HD~165141 is a K0III Ba1 star rotating much faster ($v\sin i =
14$~\kms) than normal K giants or barium stars.  As shown by de
Medeiros, Mayor \& Simon (1992), the $v\sin i$ distribution of normal
(single) K giants peaks at very low values ($v\sin i < 1$~\kms) and
decreases rapidly with increasing $v\sin i$. The median of this
distribution amounts to 2~\kms, with 97\% of the stars having $v\sin i
< 10$~\kms.

The rapid rotation of HD~165141 probably accounts for the unusually
high level of chromospheric activity exhibited by that star as
compared to other barium systems (CaII and CIV~$\lambda 1550$
emission, relatively hard X-rays detected by ROSAT). Actually, the
system has typical features of a RS~CVn system, except for the long
orbital period of 5200~d. This long orbital period makes it very
likely that wind accretion rather than RLOF is responsible for the
chemical peculiarities of this barium star. Our predictions for spin
accretion presented in Sect.~\ref{sect:spinup} may thus be applied to
the HD~165141 system, which is too wide for tidal coupling to be
responsible for the fast rotation rate, as is the case in genuine RS
CVn systems.

Adopting $\Delta L \Delta M_2^{-1} A^{-1} V_c^{-1} \sim 0.06$, as read
from Table~ \ref{table:runs}, we use $R_2 = 10$~$\Rsun$ (see below), 
$r_g = 0.3$
(e.g. Tout \& Hall 1991), $M_1 = 0.6$~$\Msun$, $M_2 = 1.5$~$\Msun$,
$V_c = 16$~\kms, $v_{\rm rot,0} = 2$~\kms\ in Eq.~(\ref{Eq:vrot}) and
find that the spin up of HD~165141 requires a fraction $\Delta M_2 /
M_{2,0} = 0.008$ to be accreted to be consistent with its high current
spin rate. The associated chemical pollution is consistent with that
star displaying only mild chemical anomalies, since ${\cal F}= 0.011$
and $\log_{10}f = 0.3$~dex.

The reason why HD~165141 seems to be the only barium star exhibiting
such a signature of spin accretion may be related to the fact that it
is also the barium star with the hottest (i.e. youngest) WD companion
(having $T_{\rm eff} \sim 35000$~K; Fekel et al. 1993).  That
situation contrasts with the other barium stars where the WD is
usually not detected by the IUE satellite, indicating that it must be
much cooler (i.e. older; see e.g. B\"ohm-Vitense et al. 1984).  The
cooling time scale of the WD companion of HD~165141 ($\tau_{\rm WD}
\sim 10^7$~y) is in fact shorter than the time spent by the present
barium star as a giant, either on the first giant branch 
(assuming in this case that it is less massive than $\sim 2 \Msun$) or in the
He clump.  It may therefore be inferred that the mass
transfer must have occurred when the barium star was already a
giant star, contrary to the situation prevailing for the majority of
barium stars that were formed as dwarfs (e.g. North \& Duquennoy 1991;
North, Berthet \& Lanz 1994).  For the dwarf stars, magnetic braking
as well as the increase of moment of inertia accompanying the radius
expansion on the giant branch will slow the rotation down (e.g. Habets
\& Zwaan 1989), thus erasing the possible spin up resulting from the
accretion process.  Magnetic braking is probably still operating in
HD~165141, on a time scale of the order of $10^8$~y if $v_{\rm rot}
\sim 14$~\kms\ [see Eq.~(2) of Habets \& Zwaan 1989]. The slowing down
of HD~165141 during the last $10^7$~y, corresponding to the time
$\tau_{\rm WD}$ elapsed since the end of the mass transfer process,
has thus been rather moderate, in agreement with the large observed
spin velocity.

The current radius of $R_2 = 10\;\Rsun$ adopted for HD~165141 in the
above analysis corresponds to
the lower limit derived
from its 35~d photometric period and its $v \sin i =
14$~\kms\ rotational velocity (Fekel et al. 1993).  
With such a radius and 
\Teff$ \sim 4700 - 5000$~K, the star must currently be either in the core
He-burning phase of evolution, or about midway on the RGB, according
to the evolutionary tracks of Schaller et al. (1993).  
At the time of accretion (i.e. about $10^7$~y ago, as given by the
cooling time of the WD companion), the stellar radius may have been 
very different from its current value, being either much smaller 
(in the case that accretion
occurred when HD~165141 was on the lower RGB, assuming it is still on
the RGB now) or much 
larger (in the case that accretion
occurred when HD~165141 was on the upper RGB, assuming that it reached
its present core He-burning stage less than $10^7$ y ago). 
It would be similar only in the case that  HD~165141 remained in the  
core He-burning phase for the last $10^7$~y (which is actually the
most probable situation, as core He-burning lasts for   
at least $2.4\;10^8$~y in solar-metallicity stars of mass $2 \Msun$ or
less). The fact that the stellar radius possibly varied 
in the course of stellar evolution following the accretion event, does 
not, however, invalidate the previous analysis based on
Eq.~(\ref{Eq:vrot}), provided that the spin velocity before accretion,
$v_{\rm rot,0}$, against which the current spin rate $v$ is compared, is 
taken equal to the typical spin rate of normal stars in
the {\it same} evolutionary phase, i.e. having the same current radius
as HD~165141. In doing so, possible radius changes 
cancel out as they affect $v$ and $v_{\rm rot,0}$ in the same way
(provided that the accretion process did not alter the gyration radius $r_g$). 

The possibility that {\it dwarf} barium stars have accreted spin
angular momentum can be investigated on the sample studied by North \&
Duquennoy (1991) and North, Berthet \& Lanz (1994). With the exception
of one star (HD~198583), which has a rotational velocity substantially
larger than F dwarfs of the same temperature, the other barium dwarfs
seem to be rotating at rates comparable to normal (non-barium) dwarfs.
However, as in the case of barium giants, no WD companion could be
detected for the barium dwarfs observed with IUE by North \& Lanz
(1991), implying ages $\ge 2\; 10^8$~y. Since such ages are larger
than the magnetic-braking time scale estimated above, the absence of
fast rotators among dwarf barium stars is not at all
unexpected. Ultraviolet data for HD~198583 would be of interest, in
order to confirm our prediction of a correlation between fast rotation
and the presence of a hot (i.e. young) WD companion.

\section{Summary}
The SPH simulations presented in this paper model the accretion of the
wind of a mass-losing star by its companion in the case where the wind
velocity is comparable to the orbital velocity. The binary motion has
then a strong influence on the accretion process, making analytic
predictions based on the Bondi-Hoyle model suspect. In fact, we find
that the mass accretion rate is at least ten times smaller than the
rate predicted by the Bondi-Hoyle prescription, amounting to about 1\%
in a binary system with $M_1 = 3$~$\Msun$, $M_2 = 1.5$~$\Msun$, $A =
3$~AU and $v_{\rm wind} = 15$~\kms (in the case of an adiabatic gas
with $\gamma=1.5$). Smaller polytropic indices lead to slightly larger
accretion rates of the order of 8\%. These values probably represent
upper limits, as the accretion rate decreases with increasing
numerical resolution.
  
Despite the fact that the efficiency of wind accretion appears to have
been overestimated in previous studies relying on the Bondi-Hoyle
prescription, wind accretion still appears efficient enough to account
for the chemical peculiarities exhibited by barium stars with orbital
periods up to about 90~y.

The accretion of transverse linear momentum, controlling the
eccentricity variation, appears to be negligible, at least in the
present simulation involving a circular orbit.

Accretion of spin angular momentum is substantial, with a
well-developed accretion disc forming in the isothermal case, contrary
to a widespread belief that wind accretion cannot lead to the
formation of such a disc. The spin up of the accreting star resulting
from wind accretion may even be considered as an important diagnostic
of this process. The mild barium star HD~165141 may be one case where
such a spin up took place recently. Such fast-rotating post-mass
transfer objects must be observed before the various braking processes
(among which magnetic braking) slow the star down again. Therefore, we
predict a correlation between WD temperature and barium star spin
rate: fast spinning barium stars are predicted to have hot WD
companions.

\section*{Acknowledgements}
T.Theuns was supported by the EEC Human Capital and Mobility Programme
under contract CT941463.

\section*{Appendix}
Following the method of Huang (1956) we derive expressions for the
change in orbital parameters due to accretion and the action of
gravitational forces. Assuming that star 1 suffers a spherically symmetric
mass loss, then conservation of linear momentum in a small interval of
time $\del t$ can be written as:

\begin{eqnarray}
\GG_1 \del t&=& \bigl[ (M_1+\del M_1)(\vv_1+\del \vv_1)-\del M_1 
                                          \vv_1\bigr] \nonumber\\
&& - \bigl[M_1 \vv_1\bigr] \label{eq:F1}\\
\GG_2 \del t&=& \bigl[ (M_2+\del M_2)(\vv_2+\del \vv_2) \bigr]\nonumber\\
&& - \bigl[ M_2\vv_2+\sum_w \del m_w \vv_w\bigr],
\label{eq:F2}
\end{eqnarray}
where the $\GG$'s refer to the gravitational force exerted by the wind
on the corresponding star and $\sum_w \del m_w \vv_w$ denotes the change
in momentum of the accreting star due to momentum accreted from the
wind. The change in orbital parameters $A$, $P$ and $e$ follows from
the change in mass, energy $E$ and angular momentum $h$ per unit reduced mass
through:
\begin{eqnarray}
\del A / A &=&\del (M_1+M_2) / (M_1+M_2) - \del E/E\label{eq:dA} \\
\del P / P &=&\del (M_1+M_2) / (M_1+M_2) - {3\over 2} \del E/E
\label{eq:dP}\\
e\del e / (1-e^2) &=& \del (M_1+M_2) / (M_1+M_2) - {1\over 2} \del E/E
\nonumber\\
&& - \del h/h\label{eq:de},
\end{eqnarray}
where $E=v_t^2/2+v_r^2/2-(M_1+M_2)/r$ and $h=r v_t$, with $r, v_r (=
\dot r$) and
$v_t$ being respectively the radius vector, the radial velocity and
the transverse velocity of the accreting star in the relative frame
where the mass-losing star is at rest. Denoting by $v_{1,y}$ and
$v_{2,y}$ the projections of the individual (inertial) velocities on
the $y$ direction transverse to the radius vector and oriented along
the orbital motion of star 1, one has $v_t = v_{1,y} - v_{2,y}$.
Orbit-averaged variations, denoted by $\langle \rangle$, can be computed from
the change in the relative transverse velocity $\del v_t$, obtained
from Eqs.~(\ref{eq:F1}) and (\ref{eq:F2}):
\begin{eqnarray}
\del v_t &=& \del v_{1,y} - \del v_{2,y}\nonumber\\
&\equiv &-F_y \del t/M_2 \nonumber\\
&=&G_{1,y} \del t/M_1 - G_{2,y}\del t/M_2 \nonumber\\
&&- \sum_w \del m_w (v_{w,y}-v_{2,y})/M_2.
\label{eq:dvt}
\end{eqnarray}
Here, as before, $G_{1,y}$ and $G_{2,y}$ are the projections of the
gravitational forces on the transverse direction. With the above
conventions, $G_{1,y}$ and $G_{2,y}$ are positive, since gas is
lagging behind the accretor with respect to its orbital motion.
Hence, the mass-losing star is accelerated by that material whereas
the accretor is decelerated. Note that $\del v_t$ refers to
the variation of $v_t$ induced by the mass transfer (and not to the variation
along the orbit due to keplerian motion).    

When computing the orbit averages $\langle \del E\rangle$ and $\langle
\del h\rangle$, it is assumed that the
mass accretion rate $\dot M_2$ and the force $F_y$  
do not depend on the phase of the binary. 
Clearly, this assumption will be poor when the eccentricity is
large, so that the relations derived here are only valid for 
systems with small eccentricities. It should also be noted that
keeping the force $F_y$ constant along the orbit is not equivalent to
Huang's parametrization in terms of constant 
$\lambda = -\del (M_2 v_t)/ V_c \del M_2  
= -v_t/V_c + F_y \del t/V_c \del M_2$. Indeed,
$\lambda$ and $F_y$ cannot both remain
constant along an eccentric orbit, since $v_t/V_c$
varies with orbital phase in an eccentric orbit. Therefore, the
relations derived here are not equivalent to Huang's. 

Averaging over the orbital period then amounts to using the following
mean values:
\begin{eqnarray}
\langle \dot r\rangle &=& 0\\
\langle \dot r^2\rangle & = & V_c^2 (1-(1-e^2)^{1/2})\\
\langle r\rangle &=& A(1+e^2/2)\\
\langle 1/r\rangle&=&1/A\\
\langle 1/r^2\rangle &=& 1/[A^2 (1-e^2)^{1/2}].
\end{eqnarray}
As an example of how the calculation is done, we will compute the
orbit-averaged change in kinetic energy per unit mass, $T={1\over 2}\,
v_t^2 + {1\over 2}\, \dot r^2$. 
Assuming $F_y$ (i.e. $\del v_t$) constant along the orbit,
the orbit-averaged differential of the first term in $T$ writes:
\begin{eqnarray}
\lan v_t\del v_t\ran &=& \lan v_t\ran\,\del v_t \nonumber\\
                     &=& - {h \over A} {F_y \del t \over M_2} 
\end{eqnarray}
where we used the average $\lan v_t\ran = h\lan 1/r \ran = h/A$. 

Analogously, we obtain for the second term:
\begin{eqnarray}
\lan \dot r \del \dot r\ran &=& \lan \dot r\, \del ({M_2\,\dot r\over
M_2}) \ran\nonumber\\
&=&\lan \dot r\,{\del (M_2\,\dot r)\over M_2} - \dot r^2\,
{\del M_2\over M_2}\ran\nonumber\\
&=&\lan \dot r\ran\,{\del (M_2\,\dot r)\over M_2} - \lan \dot
r^2\ran\,{\del M_2\over M_2}\nonumber\\
&=&- V_c^2\,(1-(1-e^2)^{1/2})\,{\del M_2\over M_2}\,
\end{eqnarray}
where we assumed $\del (M_2\,\dot r)$ to be constant over an orbital
period and used the averages in Eqs.~(27--28). The sum of these last
two results gives $\lan \del T\ran$.

Proceeding in the same spirit, we get
\begin{eqnarray}
{\lan \del\Omega\ran\over E} &=& 2\,{\del M_1+\del M_2\over M_1+M_2}\\
{\lan \del h\ran\over h} &=& -{(1 + e^2/2) \over
                              (1-e^2)^{1/2}} {F_y \del t \over V_c M_2}
\end{eqnarray}
where $\Omega$ denotes the potential energy per unit mass. Using
these orbital averages in Eqs.~(23-25) then leads to
Eqs.~(\ref{eq:dAdt}--\ref{eq:dedt}) at order $e^2$.


The change in eccentricity given by Eq.~(\ref{eq:de}) can be
understood intuitively as follows: first recall that, for a given
energy, the circular
orbit has the maximum angular momentum. It follows that increasing the
angular momentum $\del h>0$ while keeping masses and energy constant,
decreases the eccentricity, hence circularising the binary. Increasing
the energy $\del E>0$, while keeping $h$ constant, has the opposite
tendency.
\end{document}